\documentclass[aps,amsmath,twocolumn,amssymb,floatfixng,showpacs, superscriptaddress,footinbib, bibliography]{revtex4-1}

\usepackage{graphicx}
\usepackage{dcolumn}
\usepackage{multirow}
\usepackage{booktabs}
\usepackage{bm,color}
\usepackage{braket}
\usepackage{amsmath,amssymb}
\usepackage[colorlinks,linkcolor=blue,hyperindex,CJKbookmarks]{hyperref}
\usepackage{epstopdf}

\newcommand{\bgea}{\begin{equation}}
\newcommand{\enea}{\end{equation}}
\newcommand{\bea}{\begin{eqnarray}}
\newcommand{\eea}{\end{eqnarray}}

\usepackage[cmyk,dvipsnames]{xcolor}
\definecolor{darkgreen}{HTML}{009900}

\begin{document}
\title{Spin-lattice couplings in two-dimensional CrI$_3$ from first-principles study}
\author{Banasree Sadhukhan}
\email{banasree@kth.se}
\affiliation{Department of Applied Physics, School of Engineering Sciences, KTH Royal Institute of Technology, AlbaNova University Center, SE-10691 Stockholm, Sweden}
\author{Anders Bergman}
\affiliation{Department of Physics and Astronomy, Uppsala University, Box 516, SE-75120 Uppsala, Sweden}
\author{Yaroslav O. Kvashnin}
\affiliation{Department of Physics and Astronomy, Uppsala University, Box 516, SE-75120 Uppsala, Sweden}
\author{Johan Hellsvik}
\affiliation{PDC Center for High Performance Computing, KTH Royal Institute of Technology, SE-100 44 Stockholm, Sweden}
\affiliation{
Nordita, KTH Royal Institute of Technology and Stockholm University, Hannes Alfvéns väg 12, SE-106 91 Stockholm, Sweden}
\author{Anna Delin}
\affiliation{Department of Applied Physics, School of Engineering Sciences, KTH Royal Institute of Technology, AlbaNova University Center, SE-10691 Stockholm, Sweden}
\affiliation{Swedish e-Science Research Center (SeRC), KTH Royal Institute of Technology, SE-10044 Stockholm, Sweden}

\begin{abstract}
Since thermal fluctuations become more important as dimensions shrink, it is expected that low-dimensional magnets are more sensitive to lattice distortions and phonons than bulk systems are. Here we present a fully relativistic first-principles study on the spin-lattice coupling, i.e. how the magnetic interactions depend on local lattice distortions, of the prototypical two-dimensional ferromagnet CrI$_3$. We extract an effective measure of the spin-lattice coupling in CrI$_3$ which is up to ten times larger than what is found for bcc Fe. The magnetic exchange interactions, including Heisenberg and relativistic Dzyaloshinskii-Moriya interactions, are sensitive both to the in-plane motion of Cr atoms and out-of-plane motion of ligand atoms. We find that significant magnetic pair interactions change sign from ferromagnetic (FM) to anti-ferromagnetic (AFM) for atomic displacements larger than 0.16 \AA.
We explain the observed strong spin-lattice coupling by analyzing the orbital decomposition of isotropic exchange interactions, involving different crystal-field-split Cr$-3d$ orbitals. The competition between the AFM t$_{2g}$ - t$_{2g}$ and FM t$_{2g}$ - e$_{g}$ contributions depends on the bond angle formed by Cr and I atoms as well as Cr-Cr distance. In particular, if a Cr atom is displaced, the FM-AFM sign change when the I-Cr-I bond angle approaches 90$^\circ$.
The obtained spin-lattice coupling constants, along with the microscopic orbital analysis can act as a guiding principle for further studies of  the thermodynamic properties and combined magnon-phonon excitations in two-dimensional magnets.
\end{abstract}

\maketitle

\section{Introduction}
\label{intro}

Until quite recently, no magnetic atomically thin, or two-dimensional (2D), material was known to exist, despite the rapidly growing number of synthesized 2D materials. Since just a few years this situation has changed due to, e.g., the discovery of magnetic long-range order in monolayer CrI$_3$ in 2017 \cite{Huang2017}. Understanding the nature of the magnetism in these materials is connected to fundamental issues in condensed matter physics such as the relation between dimensionality, thermal fluctuations and critical behavior, and the onset of topological order in low-dimensional magnetic systems. 
In addition, 2D magnetic materials hold promise for several technological applications, e.g., transistors with magnetic functionality, high-efficiency spin filters, and ultrathin magnetic sensors \cite{Wang2018,Song2018,xing2017electric,huang2018electrical,Jimenez2020}. Unsurprisingly, these fascinating materials have quickly become a very active field of research.
%
In particular, CrI$_3$ has become somewhat of a canonical system for exploring magnetism in 2D, probably due to that it was one of the very first 2D magnets to be discovered.
In CrI$_3$, the large spin-orbit coupling (SOC) in the I-ions create a substantial magnetic anisotropy, stabilizing the 2D magnetic long-range order\,\cite{Huang2017}. 
In fact, large enough SOC may also stabilize magnetic long-range order even in atomic chains, according to some predictions\,\cite{Smogunov2008}.
In 2D CrI$_3$, the magnetic interaction between the individual Cr atoms are of superexchange type, mediated through the I-ions. The angles in the Cr-I-Cr bonds are close to 90 degrees, which implies that it is unclear from the Goodenough-Kanamori rules whether the Cr-Cr nearest-neighbor (NN) interaction will be ferromagnetic (FM) or antiferromagnetic (AFM). Instead, careful computations need to be performed. 
The magnetic Cr ions in CrI$_3$ form a honeycomb lattice, and topological edge magnons have been predicted to exist in such systems from general arguments\,\cite{Zhang2013}, a prediction which was
subsequently indirectly confirmed experimentally in CrI$_3$ \cite{Chen2018}, using inelastic neutron scattering to map out the magnon spectrum.
%
To analyze the physics behind the topological magnon gap in these systems in more detail, the magnetic interactions and magnetic excitations (magnon spectra) were recently computed with DFT+$U$ including spin-orbit coupling for bulk, 2D and strips of CrX$_3$ (X = Cl, I, Br) systems\,\cite{PhysRevB.102.115162}. These calculations show that a small topological magnon gap is formed, supporting the view that these systems are indeed topological magnetic insulators (TMI). However, the obtained gap is minute and much smaller than the experimentally observed gap. In the same work, it is speculated that the disagreement between previous calculations and experiment might originate from lattice effects. 
Supporting this view is the fact that the phonon modes involving Cr atoms in CrI$_3$ have, using DFT calculations, earlier been found to be particularly sensitive to the magnetic ordering, suggesting substantial spin–lattice and spin–phonon coupling in this system\,\cite{webster2018distinct}.
The effect of lattice vibrations on these chiral edge magnon modes was recently investigated theoretically\,\cite{Thingstad2019}, finding that lattice vibrations may weaken the topological protection, and that magnon polarons many form.
Clearly, the nature and effect of the spin-lattice couplings in  CrI$_3$ seem to be complex issues, warranting careful investigation.

In the present work, we compute the spin-lattice interactions in CrI$_3$, analysing how both the Heisenberg interaction parameters $J_{ij}$ and the Dzyaloshinskii-Moriya interaction (DMI) parameters $D_{ij}$ change when the Cr atoms are displaced from their equilibrium positions. We also present a detailed orbital analysis of the magnetic interactions using superexchange theory.

The manuscript is organized as follows:  In Sec. \ref{method} we describe the method and techniques for the coupled SLD.  We present the formalism for computing magnetic interactions within relativistic limit and give details of the performed calculations.  In Sec. \ref{result}, we present our main results,  namely the effect of lattice displacements of Cr and I atoms the on magnetic exchange interactions, DMI.  We also present an orbital analysis for deeper microscopic understanding of how the magnetism can be tailored by the lattice displacements and estimate the SLD constant for CrI$_3$ monolayer.  Section \ref{conclusion} summarizes our conclusions and gives an outlook.

\section{Theory and computational details}
\label{method}

The bilinear spin Hamiltonian $\mathcal{H}_{\rm S}$ contains Heisenberg exchange, Dzyaloshinskii-Moriya interaction (DMI) and symmetric, anisotropic interactions that in a compact form can be expressed as
\bea
\mathcal{H}_{\rm S} = -\sum_{ikj} \sum_{\{\alpha,\beta\}} e^{\alpha}_i J^{\alpha \beta}_{ij} (\{u_k^{\mu}\}) e^{\beta}_j 
\label{eq:HamS}
\eea
where $e^{\alpha}_i$ ($e^{\beta}_j$) is the $\alpha$ ($\beta$) component of the unitary vector pointing along the direction of the spin located at the site $i$ ($j$).  The exchange tensor $J_{ij}^{\alpha\beta}$ is a rank 2 tensor in spin space, with elements that in general have a dependence on the atomic displacements $\{u_k^{\mu}\}$ as well as the magnetic configuration. For clarity, the exchange tensor $J_{ij}^{\alpha\beta}$ explicitly depends on $\{u_k^{\mu}\}$. The antisymmetric part of $J^{\alpha \beta}_{ij}$ can be rewritten in terms of DMI vector,  having e.g. an $z$-component
\bea
\vec D^z_{ij} = (J^{xy}_{ij} - J^{yx}_{ij})/2. 
\label{eq:dz}
\eea
The contributions to the mixed spin-lattice Hamiltonian 
can then be obtained by expanding the bilinear magnetic Hamiltonian $\mathcal{H}_{\rm S}$ in displacement. This procedure of coupled SLD has been formulated and applied successfully to model non-relativistic exhange striction for bcc Fe \cite{PhysRevB.99.104302}. In the current manuscript, we generalize the idea by considering full exchange interaction tensors of $J^{\alpha \beta}_{ij}$ and focus on the three-body interaction
\bea
\label{eq:HamMML}
\mathcal{H}_{\rm SL}  = -\sum_{ijk} \sum_{\{\alpha,\beta\}}\Gamma_{ijk}^{\alpha\beta\mu} e_i^{\alpha} e_j^{\beta} u_k^{\mu}, 
\eea
where the coupling tensor is defined as
$\Gamma_{ijk}^{\alpha\beta\mu}=\frac{\partial J_{ij}^{\alpha\beta}}{\partial u_k^{\mu}}$.   
The so developed approach is used to calculate from microscopic origins the SLD interaction, and its effect on the magnetism, of CrI$_3$ monolayer.

The exchange interaction tensors $J_{ij}^{\alpha\beta}$ were calculated by means of magnetic force theorem, as implemented in 
the full-potential linear muffin-tin orbital-based code RSPt \cite{RSPt,  Wills2010fes}. In this approach, the exchange interactions are calculated via Green's functions within linear response theory by perturbing the spin system by deviating the initial moments ($\vec e_0$) with small angles 
\cite{antropov1997exchange,  PhysRevB.68.104436,  PhysRevB.79.045209,  secchi2015magnetic}. All components of the $J^{\alpha\beta}_{ij}$ tensor are obtained from the second order in the tilting angles. 
$\bar J$ represents the isotropic (Heisenberg) part of the interaction, and $\mid \vec D \mid $ is the magnitude of the DMI vector. 

\par The density functional calculations are performed with local spin density approximation (LSDA).  We constructed CrI$_3$ monolayer from bulk structure, with a vacuum of about 20 {\AA} added between the layers to avoid interactions between them.  Then the crystal structure was relaxed using the projector augmented wave method (PAW) \cite{PhysRevB.59.1758}, as implemented in the VASP code \cite{PhysRevB.54.11169,  kresse1996efficiency}.  The plane-wave energy cutoff was set to 370 eV with a 16$\times$16$\times$1 k-point mesh.  The exchange interaction tensors $J_{ij}^{\alpha\beta}$ were calculated using PBE \cite{PhysRevLett.77.3865} as implemented in the full-potential linear muffin-tin orbital-based code RSPt \cite{rsptweb,  wills2010full}. To calculate the spin-lattice coupling constants,  we consider 2$\times$2$\times$1 supercells in which one atom is displaced along a specific direction $\mu$. The magnetic exchange interactions $\bar{J_{ij}}$'s have been calculated within RSPt \cite{RSPt,  Wills2010fes} for both the unitcell and  2$\times$2$\times$1 supercell with 16$\times$16$\times$1 k-point mesh for which identical $\bar{J_{ij}}$'s were obtained.   In order to calculate spin-lattice coupling constants,  $\bar{J_{ij}}$ have been calculated for the same 2$\times$2$\times$1  cell but now with one atom displaced with a finite displacement $\Delta{U}$ along $\mu$ direction.  Here we consider the displacement of both the magnetic and ligand atoms along in-plane ($x$, $y$,  $xy$) and out-plane ($z$) directions, respectively, depending on the lattice geometry of the CrI$_3$ monolayer.
Here the $x$, $y$,  $xy$, $z$ represent the [100], [010], [110], [001] directions respectively. From the displaced supercell,  we calculated the spin-lattice coupling constants as given by $\Gamma_{ijk}^{\alpha\beta\mu}=\frac{\partial J_{ij}^{\alpha\beta}}{\partial u_k^{\mu}}$.  

In order to analyze the strength of the spin lattice coupling we define the quantities
\begin{eqnarray}
\Gamma_{ijk}^{\mu} = \frac{\Gamma_{xx}^{\mu}+\Gamma_{yy}^{\mu}+\Gamma_{zz}^{\mu}}{3},  \\
|\Gamma_{ijk}^{\mu}| = \sqrt {(\Gamma_{xx}^{\mu})^2+(\Gamma_{yy}^{\mu})^2   +(\Gamma_{zz}^{\mu})^2}, \\
\Gamma_{ijk} = \frac{{\Gamma_{ijk}^{\mu=x}}+{\Gamma_{ijk}^{\mu=y}}+{\Gamma_{ijk}^{\mu=z}}}{3},  \\
|\Gamma_{ijk}| = \frac{|{\Gamma_{ijk}^{\mu=x}}|+|{\Gamma_{ijk}^{\mu=y}}|+|{\Gamma_{ijk}^{\mu=z}}|}{3},  
\label{def-sld}
\end{eqnarray}
where $\Gamma_{ijk}$ and $|\Gamma_{ijk}|$ are an isotropic spin lattice coupling constant and its absolute value, respectively.

\begin{figure*} [ht] 
\includegraphics[width=1.05\textwidth,angle=0]{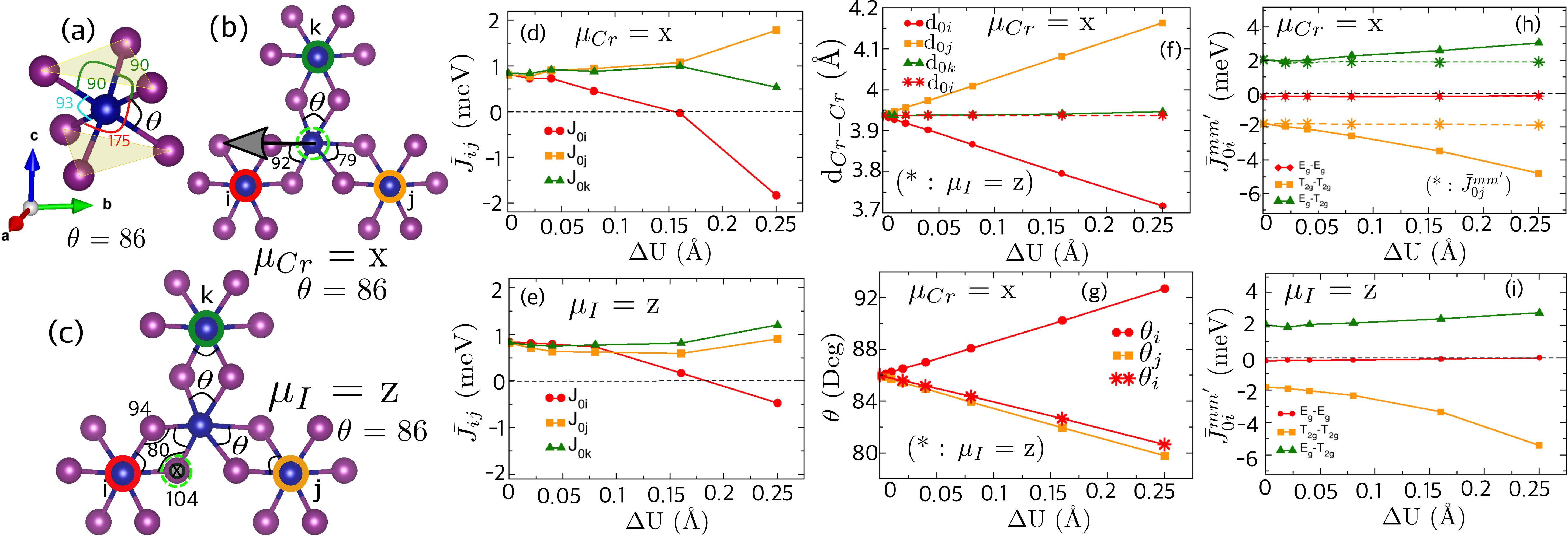} 
\caption{(a) I-Cr-I bond angles in the octahedral environment of Cr atoms.  The change in I-Cr-I bond angles of NN $i$-link,   $j$-link,  $k$-link with (b) $\mu_{Cr} = x$ and (c) $\mu_{I} = z$ respectively.  Here the displacement magnitude is chosen as $\Delta{U}=0.25$ \AA.  The green circle indicates the Cr or I atom being displaced.  Calculated isotropic exchange interaction (${\bar J_{ij}}$) with (d) $\mu_{Cr} = x$ ; (e) $\mu_{I} = z$.  The change in (f) Cr-Cr bond distance and (g) I-Cr-I bond angles for the NN $i$-link,   $j$-link,  $k$-link with $\mu_{Cr} = x$ ( $*$ corresponds to $\mu_{I} = z$).  Calculated orbital-resolved exchange interaction of the NN $i$-link ($\bar J_{0i}^{mm'}$) with displacement of (h) $\mu_{Cr} = x$ (* corresponds to the NN $j$-link ($\bar J_{0j}^{mm'}$)) and (i) $\mu_{I} = z$. }
\label{fig1} 
\end{figure*}

\section{Results} 
\label{result}

\subsection{Sensitivity of magnetism with displacement of atoms}
\par Bulk CrI$_3$ crystallizes in  a layered van der Waals material (low temperature space group R$\bar{3}$) which can easily be exfoliated to produce 2D monolayers.  The optimized in-plane lattice constant for monolayer CrI$_3$ is $a=6.817$ \AA.  Monolayer CrI$_3$ consists of the honeycomb arrangement of the Cr atoms coordinated by the six I ligands which form a distorted corner-sharing octahedral environment around each Cr atom.  The honeycomb network of Cr ions is sandwiched by two atomic planes of I atoms.  To study the effect of lattice displacements on the magnetic interactions, we consider a 2$\times$2$\times$1 supercell.  In octahedral environment,  the I-Cr-I bond angle (same plane of I atoms) within different octahedra is approximately $90^{\circ}$, whereas the I-Cr-I the band angle (different plane of I atoms) is less than $180^{\circ}$ within the same octahedra.  The  I-Cr-I bond angle for the NN links ($i$-link,   $j$-link,  $k$-link) of Cr atom connecting the I ligands of opposite planes is approximately $86^{\circ}$ (see Fig.~\ref{fig1}(a)).  

\par  Magnetism in CrI$_3$ is associated with the partially filled $d$ orbitals of Cr atom with an electronic configuration $3s^03d^3$.  In octahedral environment of Cr atom, the crystal field interaction with the I ligands results quenching of orbital moment (L = 0)  and splitting of 3$d$ orbitals into $e_g$ ($d_{x^2-y^2}, d_{z^2}$) and $t_{2g}$ ($d_{xy},  d_{yz},  d_{zx}$) manifolds.  Therefore, the three electrons occupying the $t_{2g}$ triplet makes  CrI$_3$ monolayer almost an ideal realization of a system with spin $S = 3/2$ according to Hund’s rule,  and gives an atomic magnetic moment of $\approx$ 3$\mu_B$ per Cr atom. 

\par The exchange parameters, calculated for both the unitcell and the supercell of CrI$_3$ monolayer, as a function of distance, are found to be in very good agreement with reported values \cite{PhysRevB.102.115162}. Here the isotropic part of the NN, next NN (NNN) and 3rd NN coupling are contributing in which $\bar{J_1}$ is the dominant one.  $\bar{J_1}$ consists of two competing terms originating from $e_g$ and $t_{2g}$ orbitals which are strongly hybridized with $p$ orbitals of the ligands.  This suggests a nontrivial role of the ligand states in the formation of magnetic ordering in CrI$_3$ monolayer. Later we will show a full multi-orbital analysis of superexchange mechanism as the origin of ferromagnetism in CrI$_3$ monolayer. 

\par For each Cr atom,  there exist three NN,  six NNN and again three 3rd NN.  As the isotropic part of the NN,   NNN and 3rd NN coupling are the most dominant ones, we consider 12 NN atoms in total when studying the effect of lattice displacements on magnetic ordering.  
For the undisplaced case,  all these three links have identical exchange interactions and DMI obeying the $C_3$ symmetry. Lattice displacements break the $C_3$ symmetry resulting a lower symmetry in the obtained set of $\bar{J_{ij}}$ and DMI. However, the exchange interactions $\bar{J_{ij}}$ for all NN of $i$-link,   $j$-link,  $k$-link take different values.  The same is also true for the NNN and 3rd NN $i$-link,   $j$-link,  and $k$-link exchange interactions and DMI.  

\par Figure \ref{fig1}(d)-(e) show the change in $\bar{J_{ij}}$ for NN $i$-link,   $j$-link,  $k$-link for the in-plane displacement of Cr-atom along the $x$, and out-plane displacement  of I-atom along $z$ directions respectively.  For the undisplaced case,  the I-Cr-I bond angle for the NN $i$-link,   $j$-link,  $k$-link  is $\approx$ $86^{\circ}$.  The I-Cr-I bond angle changes for the $i,j$-link,  $j,k$-link,  $k,i$-link, and $i$-link with the displacements of Cr atom long $x, y, xy $, and I atom long $z$  respectively as shown in the Fig.~\ref{fig1}(b)-(c) where the green circle indicates the displaced Cr-atom. (see Fig. \ref{fig4} in Sec. \ref{appendix} for the other two case of displacements $\mu_{Cr}=y, xy$). 
The strength of FM exchange interaction for the NN $i$-link decreases with displacements of Cr atom along $x$, while the NN $j$-link and $k$-link follow the opposite trends. The strength of the NN $j$-link and $i$-link exchange interaction decrease, when the displacements are applied along $y$ and $xy$ directions, respectively. The FM to AFM sign change for a particular NN  link ($\bar{J}_1$) occur for  $\mu_{Cr}^{xy} \ge  0.12$ \AA,  $\mu_{Cr}^{x/y} \ge  0.16$ \AA,  $\mu_{I}^{z} \ge  0.18$ \AA\ which is 1.76\%,  2.35\%,  2.64\% of the lattice constant, respectively.  

\par The transition from the FM to AFM coupling is controlled either by the change in the I-Cr-I bond angle or by the distance between the corresponding Cr atoms. Since both parameters change when the atoms are displaced, at this stage it is hard to identify the main driving force of the sign flip of the $\bar J_{ij}$.

\par  In order to elucidate this,   we calculated the Cr-Cr bond length d$_{Cr-Cr}$ and I-Cr-I bond angle for different displacements for the in-plane motion of Cr atom ($\mu_{Cr}=x$) and out-of-plane motion of ligand atom ($\mu_{I}=z$) as shown in Fig.~\ref{fig1} (f)-(g) (see also Fig \ref{fig4} in Sec. \ref{appendix} for the other two cases $\mu_{Cr}=y, xy$). In the undisplaced case, the $d_{\text{Cr-Cr}}$ for the NN links is 3.938~\AA~which decreases with the displacement of Cr atom and switches to AFM, when it reaches the value of 3.796~\AA. 
The bond angle increases with the displacement of Cr atom and the NN coupling is FM until the I-Cr-I angle reaches $ 90^{\circ}$
(see $i$-link $J_{0i}$ for $\mu_{Cr}=x$ in Fig.~\ref{fig1} (g)). The I-Cr-I bond angles of NN-links increase and decrease for the in-plane displacement of Cr atom when the bond length decreases and increases respectively for the corresponding NN-links upon displacement ($\mu_{Cr}=x, y, xy$). The corresponding NN coupling became AFM when I-Cr-I bond angle exceeds $ 90^{\circ}$ for the displacement of Cr atom ($\mu_{Cr}=x, y, xy$, and, see also Fig. \ref{fig4} in Sec. \ref{appendix}). While the Cr displacement entails the changes of both the Cr-Cr distance and the bond angles, the movement of ligand atom along ``$z$" direction only affect the latter. In this case, we also find that the displacement of I atom affects the interaction between the Cr atoms, it is linked to, and can also change the sign of the NN coupling to AFM for sufficiently large position shift ($\Delta U = 0.25$~\AA).

\par The strength of the exchange interactions reduce to NNN-links and always have the FM sign with the displacements of Cr atom to any directions (see Fig. \ref{fig5}(a)-(e) in Sec.~\ref{appendix}). The angles between both the NNN-links and 3rd NN are 120$^{\circ}$ and change according to the displacement directions.  The $\bar J_{ij}$ for the NNN $i$-link and $j$-link follow the almost opposite trends with displacements whereas the it is unaffected for the $k$-link with the $x$-displacement of Cr atom.  This is due to fact that the bond angles for NNN $i,k$-link and $j,k$-link pairs remain same whereas it changes for the other NNN-link pairs with displacements.  

\par The strength of $\bar{J_{ij}}$ for 3rd NN-links with regard to NN-links reduces further.  The AFM $\bar J_{ij}$ for the 3rd NN-links decreases with the displacements of Cr atom and one of the link ($k$-link) flips its sign at $\mu = 0.25$ \AA ($\mu=x$) following the angle rules of 3rd NN-link (see Fig. \ref{fig5}(f)-(j) in Sec.~\ref{appendix}). The 3rd NN links need much larger displacements with regard to NN links to change its sign from FM to AFM with $\mu_{Cr}^{x} \geq  0.21$ \AA\ which is 3.1\% of the lattice constant for the CrI$_3$ monolayer.

\subsection{Orbital resolved magnetism}

\begin{figure} [h!] 
\includegraphics[width=0.499\textwidth,angle=0]{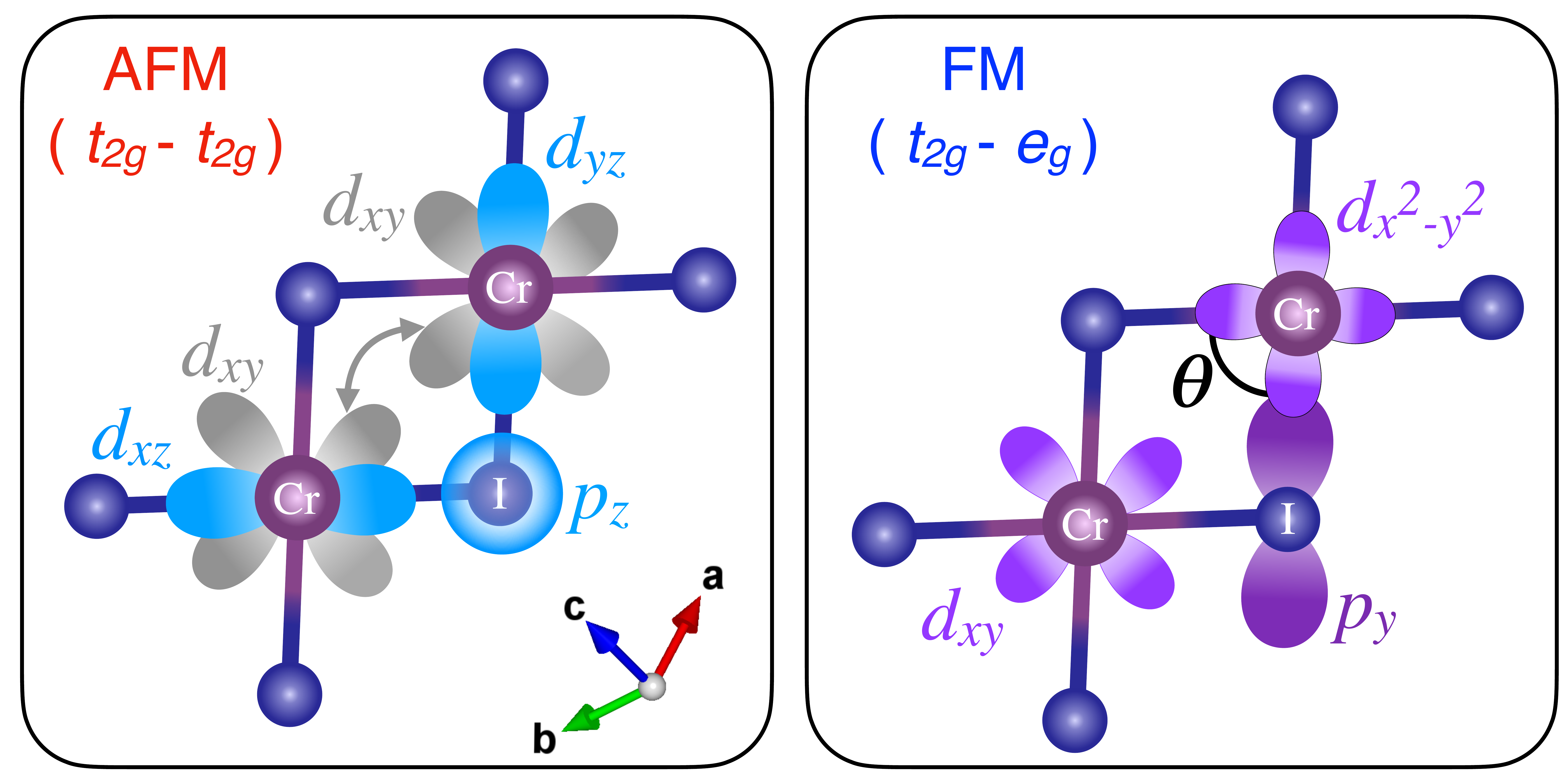} 
\caption{Schematic representation of various processes giving rise to the NN exchange interaction. The color of the orbital denotes the exchange path it belongs to. Gray arrows indicate the direct hopping process between two Cr-$t_{2g}$ orbitals.}
\label{fig-orb} 
\end{figure}

In order to get an insight into the physical origin behind the observed changes in the exchange interactions $\bar{J_{ij}}$, we performed their orbital-by-orbital decomposition. 

\par The sign of the superexchange interaction depends on the symmetry of electron orbitals arising from the crystal field the Cr atom experiences. In case of CrI$_3$, the Cr atoms are surrounded by 6 iodine atoms forming a nearly ideal octahedron. For simplicity, we assign the Cr-$d$ orbitals to $t_{2g}$ and $e_g$ subsets, which would arise in the ideal case. Although the nominal occupation of the $3d$-states of Cr$^{3+}$ ions should be roughly three, resulting in a half-filled $t_{2g}$ and empty $e_g$ shells, the electronic structure calculations reveal a different situation \cite{PhysRevB.99.104432,Kashin_2020}. There is a finite occupation of the $e_g$ orbitals, which have lobes pointing directly towards iodine atoms thus forming strong $\sigma$ bonds. As in previous reports \cite{PhysRevB.99.104432,Kashin_2020,PhysRevB.102.115162}, we find that there are two main competing contributions to the long-range magnetic ordering in CrI$_3$ which originate from the $t_{2g}-t_{2g}$ and $t_{2g}-e_{g}$ interacting orbital channels.  The total exchange interaction in CrI$_3$ from the multi-orbital approach can be presented as a sum of two contributions: $\bar{J_{ij}} = \bar{J_{ij}}^{t_{2g}-t_{2g}} + \bar{J_{ij}}^{t_{2g}-e_{g}}$. The coupling between nominally empty $e_g$ orbitals is negligibly small. The ${t_{2g}-e_{g}}$ contribution is FM involving the transitions between half-filled $t_{2g}$ and empty $e_{g}$ orbitals via a single intermediate I-$p$ orbital.

\begin{figure} [ht] 
\centering
\includegraphics[width=0.4\textwidth,angle=0]{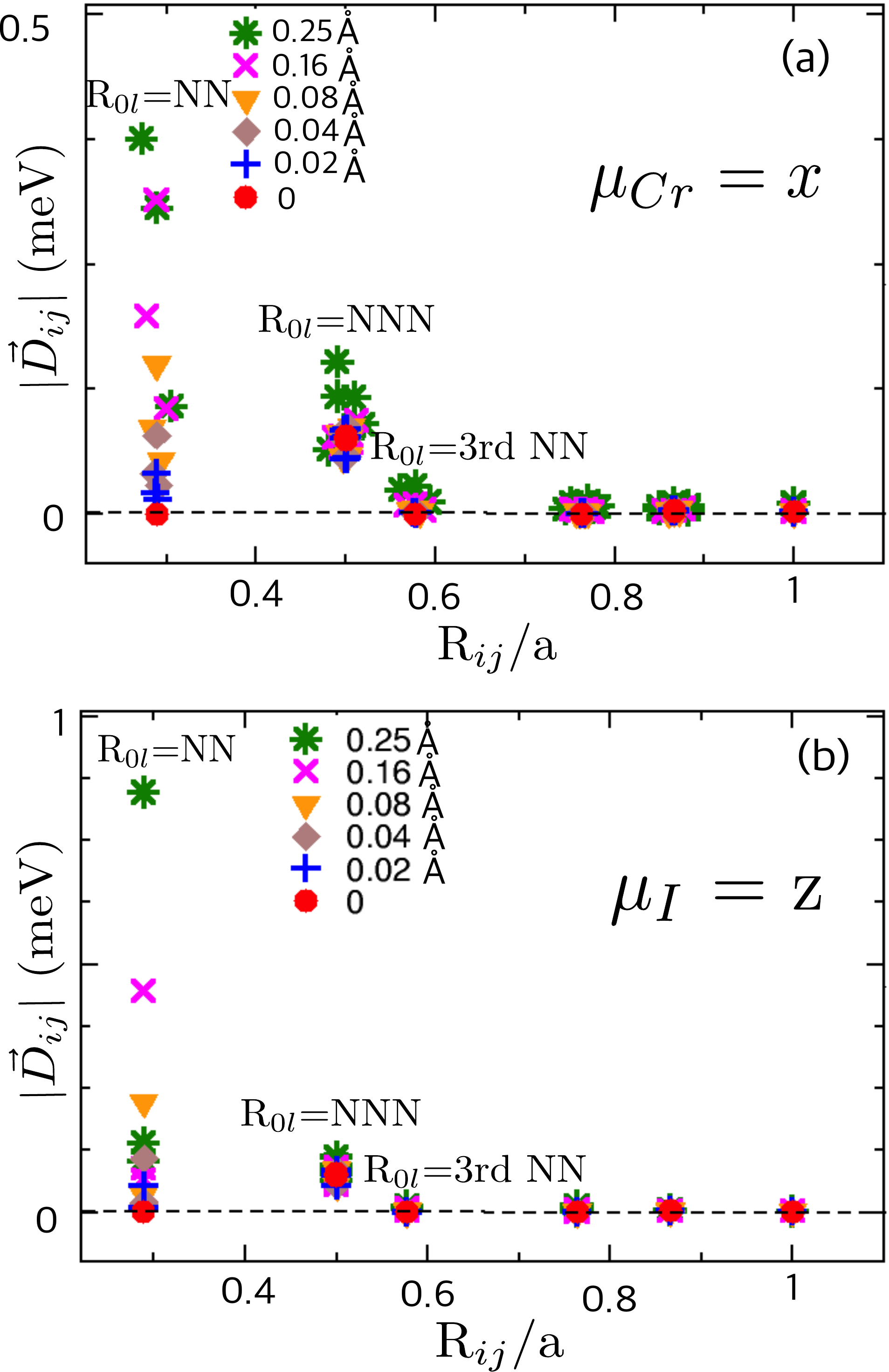} 
\caption{Dzyaloshinskii-Moriya Interactions (DMI) terms for the $i$-link,   $j$-link,  $k$-link with (a) $\mu_{Cr}$ = $x$ and (b)  $\mu_{I}$ = $z$.}
\label{fig2} 
\end{figure}

\par Figure \ref{fig1}(h) shows the orbital decomposition of $\bar{J_{ij}}$ for the NN $i$-link for the displacement of Cr atom along $\mu = x$.  
For undistorted CrI$_3$,  the FM coupling that is due to $t_{2g}-e_{g}$ hybridization  dominates over the AFM $t_{2g}-t_{2g}$ term, which leads to the stabilization of the FM ground state. If we displace the Cr atoms along $\mu = x$,  both interacting channels tend to increase with the magnitude of displacements, but the growing rate is higher for the $t_{2g}-t_{2g}$ term than for its $t_{2g}-e_{g}$ counterpart. At the displacement magnitude $|\Delta U_{\mathrm{Cr}}| = 0.01$ \AA, the  AFM $t_{2g}-t_{2g}$ term starts to dominate, and the sign of the total NN coupling flips. 
We note that the orbital decomposed $\bar{J_{0i}}$ does not change much for the displacement of Cr atom along $\mu = y$. 

The out-of-plane displacement of iodine ($\mu_{I} = z$), which brings I atom closer to the Cr-Cr plane, leads to the drastic enhancement of the AFM $t_{2g}-t_{2g}$ contribution to the NN coupling, while the FM contribution grows slower, as shown in Fig.~\ref{fig1}(i). This only holds true for the Cr-Cr bond, which is mediated by the displaced iodine atom. The other two NN couplings remain unaffected, since their exchange paths are intact. 

Fig.~\ref{fig-orb} contains a schematic picture of the orbital contributions to the NN exchange coupling. The $t_{2g}$-$t_{2g}$ contribution can be characterized by two distinct processes, which both lead to AFM interaction, as was demonstrated in Ref.~\cite{PhysRevLett.127.037204} for monolayer of CrCl$_3$. One of them is the Anderson’s superexchange, which involves the $t_{2g}$ orbitals having distinct symmetries on the two Cr atoms, mediated by the hybridization with $I-p$ states. Another one is a \textit{direct} kinetic exchange between the $t_{2g}$ orbitals of the same symmetry, whose lobes point towards each other. The FM and AFM contributions are of comparable size and since the I-Cr-I bond angle is close to 90$^{\circ}$, the balance between them can easily be altered by lattice distortions.

%
\begin{figure*} [ht] 
\includegraphics[width=1.05\textwidth,angle=0]{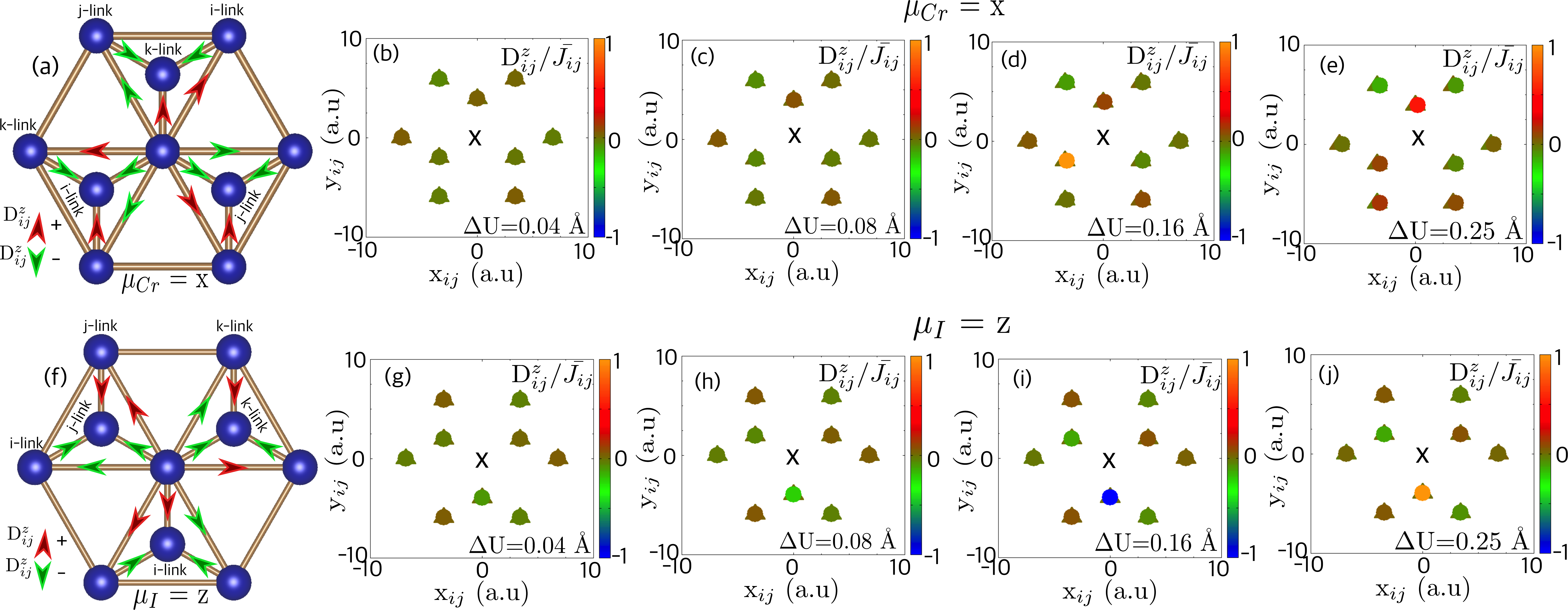} 
\caption{ The Dzyaloshinskii-Moriya interaction (DMI) for the NN and NNN links (a) $\mu_{Cr} = x$  and (f) $\mu_{I} = z$ in which the color corresponds to the sign of the z component of the DMI vectors. The ratio of the DMI and Heisenberg interaction for the NN and NNN links as a function of distance along x-axis ($x_{ij}$) and y-axis ($y_{ij}$) with (b)-(e) $\mu_{Cr} = x$  and (g)-(j) $\mu_{I} = z$.  The $\times$ represents the position of the origin i.e reference Cr atom (i) and the interactions are taken between i and its neighbors,  j with $r_{ij} = (x_{ij}, y_{ij})$.  The strength of the ratio is given by the color of the symbols where $\Delta$ and $\circ$ represent the undisplaced and displaced case respectively in each figure. }
\label{fig8} 
\end{figure*}

In the case of Cr motion, the bond acquiring the shortest Cr-Cr distance is characterized by the strongest enhancement of the constituent orbital interactions. 
Both FM and AFM contributions are expected to increase, primarily because of increased Cr-I orbital overlaps, which results in larger hopping amplitudes, generating the superexchange. At the same time, the $t_{2g}-t_{2g}$ channel also has a contribution stemming from the kinetic exchange between the orbitals, pointing directly towards each other (see Fig.~\ref{fig-orb}). This latter contribution depends primarily on the Cr-Cr distance, and is less subject to the changes in the I-Cr-I bond angle. In this case it is also expected to grow as a function of displacement amplitude and is likely to play the dominant role in the enhancement of the $t_{2g}-t_{2g}$ term, similarly to the case of the uniform strain in monolayered CrCl$_3$\cite{PhysRevLett.127.037204}.

Here, similarly to the case of Cr displacement, the increase in the orbital overlap leads to the growth of AFM $t_{2g}-t_{2g}$ term. However, the case of $\mu_I=z$ is different from $\mu_{\text{{Cr}}}=x$, since the direct kinetic exchange path is not affected by the changing position of I atom. Instead, the superexchange processes are altered. Inspection of Fig.~\ref{fig-orb} suggests that by making the I-Cr-I angle ($\theta$) smaller, one increases the overlap between the $t_{2g}$ orbitals involved in AFM superexchange (shown with blue color). At the same time, the displacement $\mu_I=z$ (along "c" direction) does not substantially increase the overlap between Cr-$e_g$ and I-$p$ states. As a result, the AFM contribution experiences a much stronger increase and thus starts to dominate at large displacement amplitudes.

\par To connect the lattice displacements where we see significant effects from the spin-lattice coupling  with a experimentally  measurable quantity,  we  have calculated the mean square displacement (MSD) at different temperatures using VASP in combination with PHONOPY \cite{PhysRevB.59.1758, PhysRevB.54.11169,  kresse1996efficiency, togo2015first,  PhysRevB.86.214301,  deringer2014ab} 
(see Fig. \ref{fig6} in Sec. \ref{appendix}). The sign change from FM to AFM for individual NN couplings occur when $\mu_{Cr}^{xy} \geq  0.12$ \AA\, $\mu_{Cr}^{x/y} \geq  0.16$ \AA, or  $\mu_{I}^{z} \geq 0.18$ \AA\  which correspond to the MSD at 132 K, 175 K, and 197 K respectively. All structures are in the low temperature rhombohedral phase which is consistent with earlier reports \cite{mcguire2015coupling}.

\subsection{Effect on Dzyaloshinskii-Moriya interactions} CrI$_3$ monolayer is reported as a topological magnetic insulator. The DMI vectors ideally lie in the plane of Cr network due to the symmetry, but a small component of the z-component of the DMI vector ($D^z$) is responsible for the opening of the topological gap at the K point. Therefore, any change in the magnetic interactions,  specially,  DMI affect the topology of magnon.  Figure \ref{fig2} depicts the DMI  as a function of distance at different displacements of Cr atom and I atom (see also Figs. \ref{fig7} and \ref{fig9} in Sec. \ref{appendix} for more details).  The NN $i$-link,   $j$-link,  $k$-link DMI are forbidden by symmetry of the systems.  Only the NNN DMI's are finite obeying the $C_3$ symmetry of CrI$_3$ monolayer and matches well in the reported values \cite{PhysRevB.102.115162}.  The $C_3$ symmetry is broken with displacements and  DMI terms for NN,  NNN and 3rd NN $i$-link,   $j$-link,  $k$-link are contributing.  The NN DMI terms increase with displacements and become dominating when ground state changes FM-AFM ordering in $\bar{J_1}$.  The dominating DMI increases almost $\sim 6$ times with regard to the undisplaced case at $\Delta{U} = 0.25$ \AA. The effect on DMI observed for the displacements of ligand atoms in coupled SLD is almost double compared to magnetic atoms.

To study the effect on DMI with the displacement of atoms in coupled SLD which in turn affect the topology of the system,  we calculate the ratio of the DMI and Heisenberg interaction ($D_{ij}^z/\bar{J_{ij}}$) for the NN and NNN links for both in-plane motion of Cr-atom and out-plane motion of I-atom as shown in Fig. \ref{fig8}. The strength of the ratio determine the domination of the DMI over the Heisenberg interaction which is the controlling parameter for magnonic topological transport.  Figures \ref{fig8}(a) and \ref{fig8}(f) depict the directions of the z-component of  DMI vectors within a Cr sublattice in coupled SLD where both the NN and NNN links are contributing.  The direction of the DMI vectors for the NN and NNN links in two adjacent Cr sublattices are opposite with regard to the other sublattice.  The DMI ratio $D_{ij}^z/\bar{J_{ij}}$  increases with increasing the displacements of Cr atom and reach a high value of 1.6 for the NN $i$-links with $\Delta{U} = 0.16$ \AA.  It is because of the $\bar{J_{0i}}$ switches from FM to AFM.  At  $\Delta{U} = 0.25$ \AA, $D_{ij}^z/\bar{J_{ij}}$  decreases (0.1) because the high value of both the DMI and Heisenberg interaction whereas the NN $k$-links reach a higher value of 0.6 determined by the coupled effect of both the interactions.

The sensitivity of the DMI ratio is larger for the displacements of ligands than magnetic atoms.  The DMI ratio $D_{ij}^z/\bar{J_{ij}}$ got a high value of 1.5 for the NN i-link at $\Delta{U} = 0.25$ \AA.  The microscopic analysis presented here shows that the DMI for some pairs is high in magnitude compared to the Heisenberg exchange in coupled SLD. 

\begin{figure} [ht] 
\includegraphics[width=0.4\textwidth,angle=0]{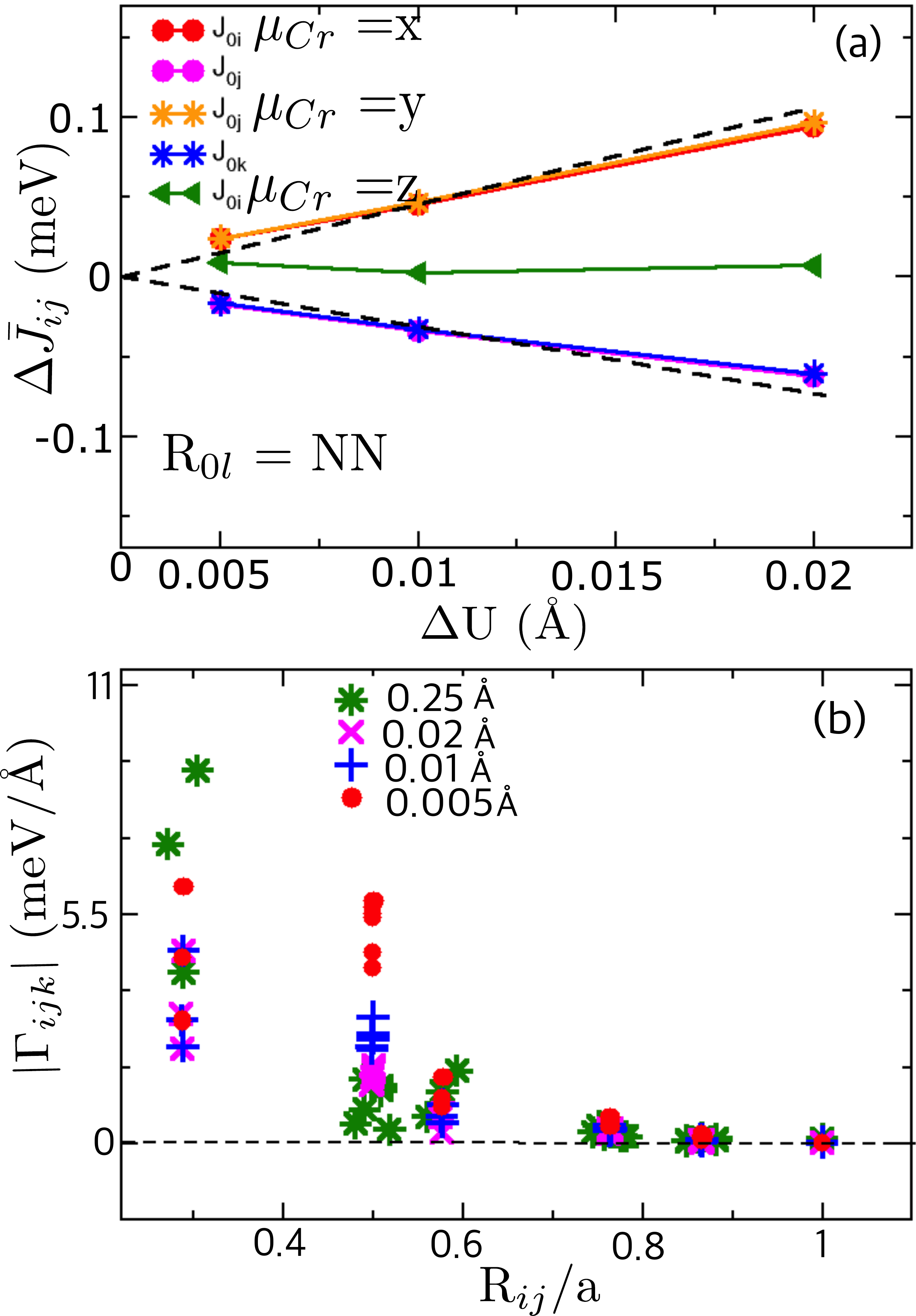} 
\caption{(a) Variation of the  isotropic exchange interaction ($\bar{J_{ij}}$) with the displacements of Cr-atom in both in-plane and out-plane directions for NN. (b) Absolute value of the isotropic spin lattice coupling constants for the $i$-link,   $j$-link, and $k$-link as a function of the distance with displacements.}
\label{fig3} 
\end{figure}

\subsection{Spin-lattice coupling:} The variation of the  isotropic exchange interaction ($\Delta{\bar{J_{ij}}}$) for NN $i$-link,   $j$-link,  $k$-link with the displacements of Cr-atom in both in-plane and out-plane directions are shown in Fig.~\ref{fig3}(a).  The system is sensitive only to the in-plane motion of the Cr-atom and out-plane motion of halide atom.  The response in the $\Delta{ \bar{J_{ij}}}$ for the $i$-link,   $j$-link,  $k$-link depends on the change in the I-Cr-I bond angle of the corresponding links.  The response of  $\Delta{ \bar{J_{ij}}} \sim \Delta{U}$ reduces to ($\sim \frac{1}{10}$) for the NNN links (see Fig.~\ref{fig10}(a) in Sec.\ref{appendix}). The sensitivity of CrI$_3$ monolayer in the linear region of coupled SLD i.e.  $\Delta{ \bar{J_{ij}}} \sim \Delta{U}$ is up to $|\mu|$ =  0.02 \AA~and beyond that limit, it falls into the non-linear region. The dotted (black) line in Fig.~\ref{fig3}(a) indicates the ideal variation of $\Delta{ \bar{J_{ij}}} \sim \Delta{U}$ which is $< {3 \%}$ offset limit of the actual response. The absolute values of the isotropic spin lattice coupling (SLC) constants for the $i$-link,   $j$-link,  $k$-link as a function of the distance with displacements are depicted in Fig.~\ref{fig3}(b). The SLC constants remain almost constants with displacements within the linear regime of coupled SLD (though it deflects slightly for smaller displacement $|\mu|$ =  0.005 \AA), and suddenly increase when they fall into the non-linear regime of coupled SLD.

The strength of the spin-lattice coupling in CrI${_3}$ can be contrasted to the strength of the exchange striction in bcc Fe \cite{PhysRevB.99.104302}. For bcc Fe, the ratio of the exchange striction coupling and the Heisenberg exchange is $|\Gamma_1|/|J_1| = 0.641$ {\AA}$^{-1}$ and $|\Gamma_2|/|J_2|= 0.481$ {\AA}$^{-1}$ for the nearest and next nearest neighbor bonds respectively. For CrI${_3}$, the corresponding ratios are $|\Gamma_1|/|J_1| = 6.97$ {\AA}$^{-1}$ and $|\Gamma_2|/|J_2|= 2.97$ {\AA}$^{-1}$, i.e. for the nearest neighbor (next nearest neighbor) bonds the relative strength of the spin-lattice coupling is a factor $\sim 10$ ($\sim 6$) stronger in CrI${_3}$ than in bcc Fe.

\section{Conclusion} 
\label{conclusion} From fully-relativistic calculations of the magnetic exchange interactions, including Heisenberg and DMI, when considering finite displacements, we have found that the spin-lattice coupling is significant in CrI$_3$. In particular it has been found that dominating exchange interactions can change sign from FM to AFM coupling when the atomic distance between neighbouring atoms increases. A microscopic explanation for the strong spin-lattice coupling based on orbital decomposition has been presented where the angle formed by I-Cr-I, and the Cr-Cr bond distance affect the magnetic interaction significantly. 
To this end, we argue that it is not enough to consider only the isotropic exchange and anisotropic DMI to study the effect of lattice displacements during thermal excitations.  It needs a orbital resolved analysis of the exchange interactions to have a complete picture from microscopic origin.  For comparison with three-dimensional ferromagnets we extract an effective measure of the SLC constants which is ten times larger for CrI$_3$ than for bcc Fe. 
The strong spin-lattice coupling in CrI$_3$ and related two-dimensional magnets are expected to play a significant role for the existence of topological magnons in these systems and we suggest that coupled spin-lattice dynamics is a suitable tool for investigating this further.

\begin{acknowledgments}
%
A.D. acknowledges financial support from Vetenskapsrådet (grant numbers VR 2015-04608, VR 2016-05980 and VR 2019-05304), and the Knut and Alice Wallenberg foundation (grant number 2018.0060).
Y.O.K. acknowledges the financial support from the Swedish Research Council (VR) under the project No. 2019-03569.
J.H. acknowledges financial support from the Swedish Research Council (VR) (neutron project grant BIFROST, Dnr. 2016-06955).
A.B. acknowledges eSSENCE.
The computations were enabled by resources provided by the Swedish National Infrastructure for Computing (SNIC) at PDC and NSC, partially funded by the Swedish Research Council through grant agreement no. 2018-05973.
\end{acknowledgments}

\bibliography{cri3-sld}{}

\section{Appendix}
\label{appendix} 

\begin{figure*} [ht] 
\centering
\includegraphics[width=0.99\textwidth,angle=0]{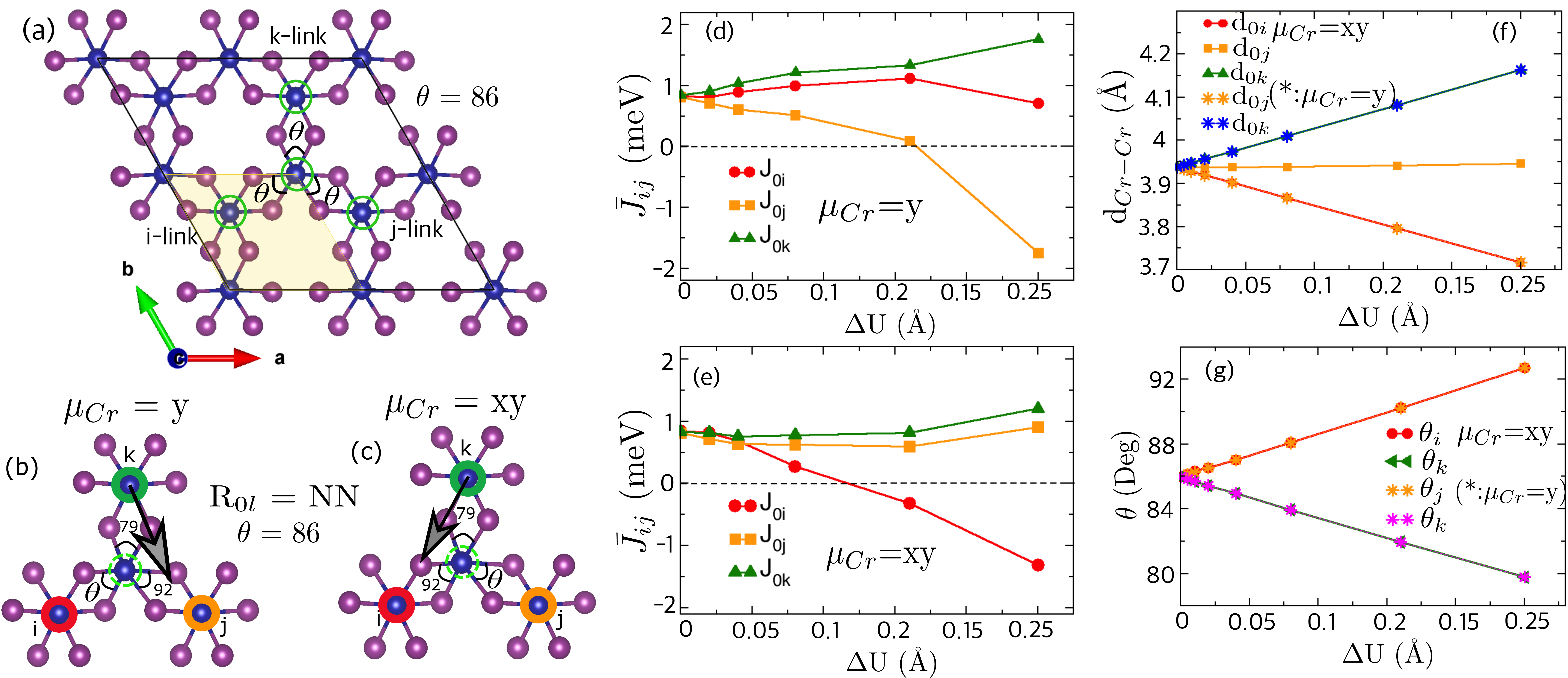} 
\caption{(a) 2$\times$2$\times$1 supercell in the $ab$-plane of CrI$_3$.  The shaded region in the supercell  corresponds to the unit cell.  The change in I-Cr-I bond angles of NN $i$-link,   $j$-link,  $k$-link with displacements along (b) $\mu_{Cr} = y$ and (c) $\mu_{Cr} = xy$ directions respectively. Here the displacement magnitude is chosen as $\Delta{U}=0.25$ \AA.  The green circle indicates the Cr atom being displaced.  Calculated isotropic exchange interaction (${\bar J_{ij}}$) with (d) $\mu_{Cr} = y$ ; (e) $\mu_{Cr} = xy$.  The change in (f) Cr-Cr bond distance and (g) I-Cr-I bond angles for the NN $i$-link,   $j$-link,  $k$-link with $\mu_{Cr} = xy$ ( $*$ corresponds to $\mu_{Cr} = y$). }
\label{fig4} 
\end{figure*}
\begin{figure*} [ht] 
\centering
\includegraphics[width=0.99\textwidth,angle=0]{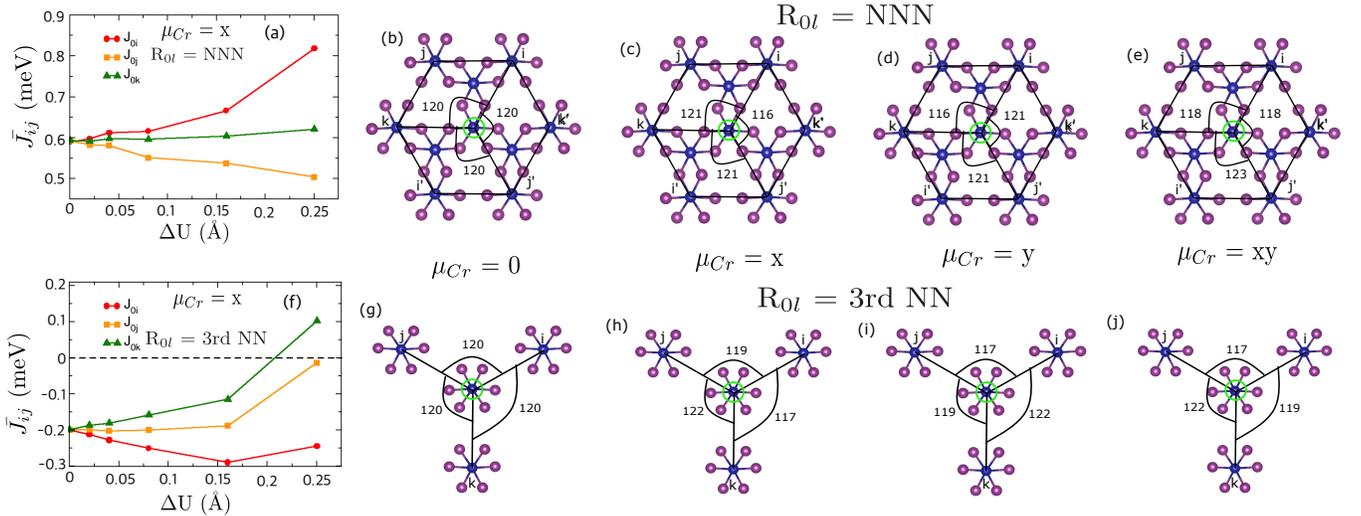} 
\caption{ The exchange interactions and bond angle for (a)-(e) NNN and (f)-(j) 3rd NN $i$-link,   $j$-link,  $k$-link without and with displacement of Cr-atom along $\mu_{Cr} = x,  y,  xy$ directions respectively .  The displacement magnitude to denote the bond angle in the NNN and 3rd NN links is chosen as $\Delta{U}$ = 0.25 \AA.  The green circle indicates the displaced Cr-atom.  }
\label{fig5} 
\end{figure*}
\begin{figure*} [ht] 
\centering
\includegraphics[width=0.4\textwidth,angle=0]{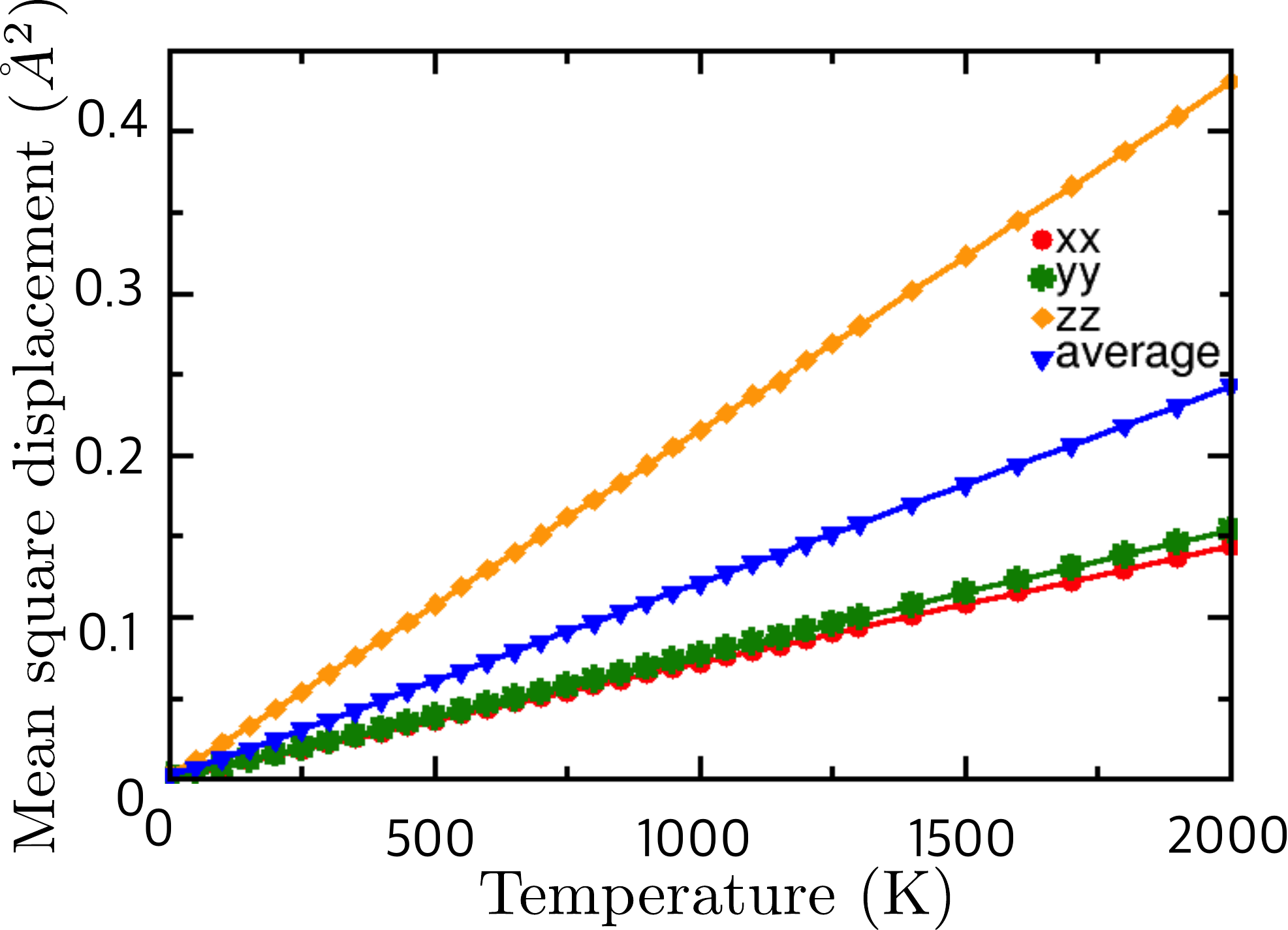} 
\caption{Calculated mean square displacements of CrI$_3$ monolayer with the temperature.}
\label{fig6} 
\end{figure*}

Figure \ref{fig4}(a) shows the geometry of the undisplaced 2$\times$2$\times$1 supercell, and Fig. \ref{fig4}(b)-(g) show the change in isotropic magnetic exchange interactions, bond distances and bond angles with the displacement along the other two in-plane motion of Cr atom $\mu=y, xy$ directions respectively. The strength of the Heisenberg exchange interactions and bond angles with the in-plane displacement of Cr atom ($\mu_{Cr} = x,  y,  xy$) for NNN and 3rd NN are shown in Fig. \ref{fig5}.  The mean square displacement of atoms as a function of temperature is shown in Fig. \ref{fig6}. 

\begin{figure*} [ht] 
\centering
\includegraphics[width=0.7\textwidth,angle=0]{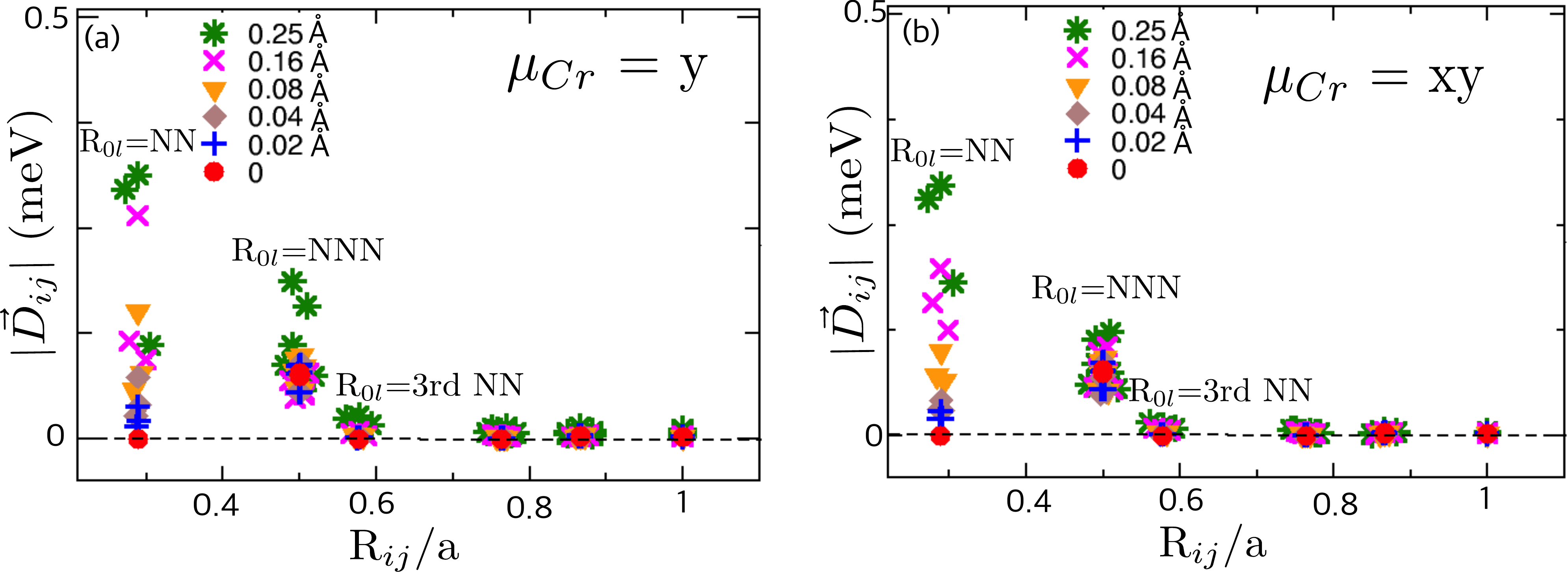} 
\caption{Dzyaloshinskii-Moriya interactions (DMI) for the $i$-link,   $j$-link,  $k$-link with (a) $\mu_{Cr}$ = $y$ and (b)  $\mu_{Cr}$ = $xy$.}
\label{fig7} 
\end{figure*}

\begin{figure*} [ht] 
\centering
\includegraphics[width=0.8\textwidth,angle=0]{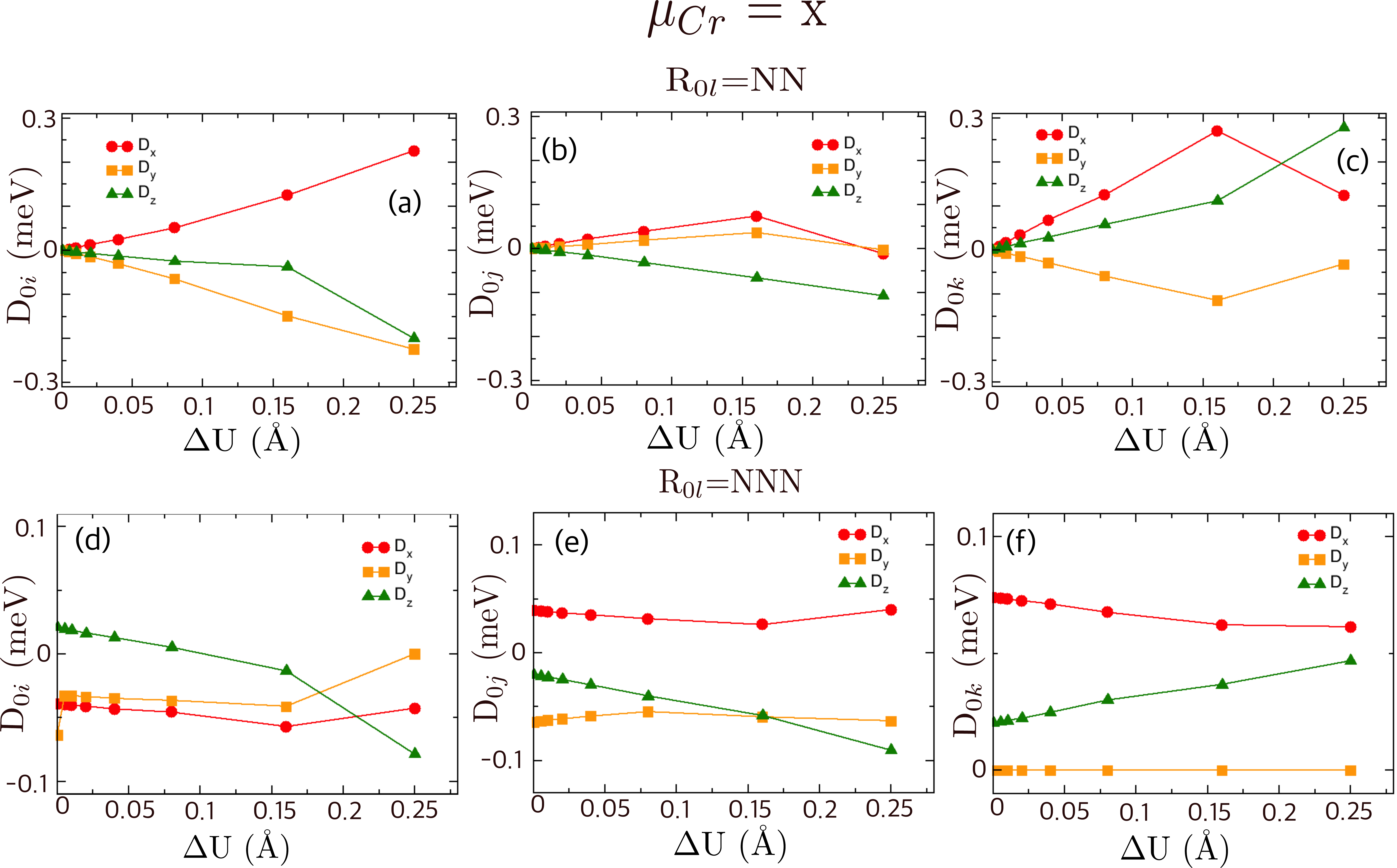} 
\caption{The components of the DMI for (a)-(c) NN links and (d)-(f) NNN links with $\mu_{Cr}$ = $x$. }
\label{fig9} 
\end{figure*}

\begin{figure*} [ht] 
\includegraphics[width=0.7\textwidth,angle=0]{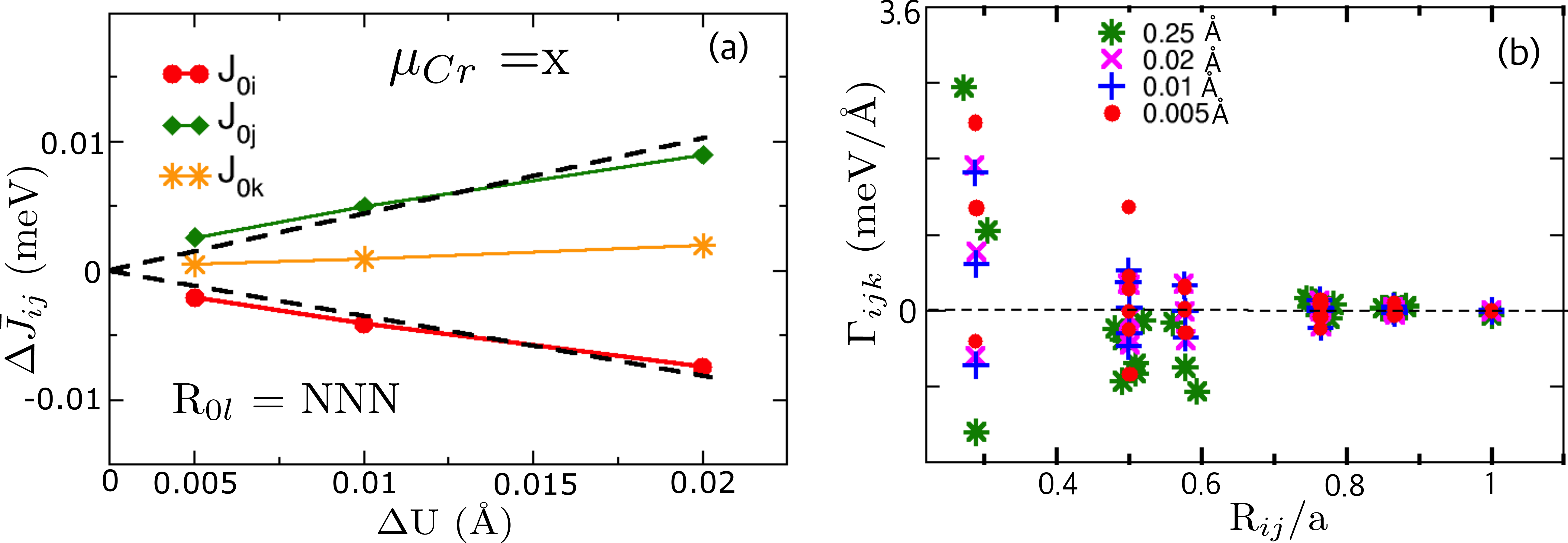} 
\caption{ (a) Variation of the  isotropic exchange interaction ($\bar{J_{ij}}$) with the displacements of Cr-atom for the NNN. (b) Isotropic spin lattice coupling constants for the $i$-link,   $j$-link,  $k$-link as a function of the distance with displacements.}
\label{fig10} 
\end{figure*}

Figure \ref{fig7} shows the change in the Dzyaloshinskii-Moriya interaction (DMI) for $i,j,j$-links of NN, NNN and 3rd NN with displacement of Cr atom along $\mu_{Cr} = y,  xy$. To analyse, how the components of DMI ($D_x, D_y, D_z$) behave with the displacements, we calculated the $D_x, D_y, D_z$ for NN and NNN $i,j,j$-links with $\mu_{Cr} = x$ as shown in Fig. \ref{fig9}(a)-(f). Figure \ref{fig10} (a)-(b) show the linear regime of coupled spin-lattice dynamics for NNN and corresponding isotropic spin-lattice coupling constants with displacements respectively. 

\end{document}